\documentclass[preprint,amsmath,amssymb,prb]{revtex4}

\usepackage{graphicx}
\usepackage{epsfig}

\begin{document}
\title{Playing Quantum Physics Jeopardy with zero-energy eigenstates}
\author{L. P. Gilbert}
\affiliation{Department of Physics, Davidson College, Davidson, North
Carolina 28035, USA}

\author{M. Belloni}
\email{mabelloni@davidson.edu}
\affiliation{Department of Physics, Davidson College, Davidson, North Carolina
28035, USA}

\author{M. A. Doncheski}
\affiliation{Department of Physics, The Pennsylvania State
University, Mont Alto, Pennsylvania 17237, USA}

\author{R. W. Robinett}
\affiliation{Department of Physics, The Pennsylvania State
University, University Park, Pennsylvania 16802, USA}

\begin{abstract}
We describe an example of an exact, quantitative
Jeopardy-type quantum mechanics problem. This problem type is based on
the conditions in one-dimensional quantum systems that allow an
energy eigenstate for the infinite square well to have zero curvature
and zero energy when suitable Dirac delta functions are added. This condition
and its solution are not often discussed in quantum mechanics
texts and have interesting pedagogical consequences.
\end{abstract}


\maketitle

\section{Introduction}
\label{sec:intro}

Students often work backward when problem solving in their
introductory physics courses. This method typically entails using
the information in the problem and if available, the answer in the
back of the book to work backward to determine which equation to
use.\cite{larkin} Students using this novice problem-solving
approach have been shown to lack the conceptual foundation of more
advanced problem solvers.\cite{chi} In more advanced courses such as
quantum mechanics, students still apply novice problem-solving
approaches and studies have shown that students' conceptual
understanding on all levels is lacking.\cite{rick_qmvi}

Van Heuvelen and
Maloney\cite{maloney} have described a problem type that encourages a more
conceptual problem-solving strategy called a ``working backward'' or
Jeopardy problem. These problems begin with the answer (an equation,
a diagram or a graph, or a simulation) and ask students to work back
toward the question, much like the game show Jeopardy.

In Quantum Physics Jeopardy, students are given an energy eigenstate
(the answer) and are asked to find the potential energy function
(the question) that yields this eigenstate. This problem elucidates
the connection between the form of the energy eigenstate and the
potential energy function in one-dimensional quantum
systems.\cite{taylor} The technique is similar to inverse problems
in quarkonium spectroscopy where constraints on the binding
potential from the bound-state energies are
obtained.\cite{quarkonium} Applications of inverse methods to other
fields, such as medical imaging\cite{saito,blackledge} (for example,
CT scans) and geophysics\cite{jacobsen} are even more familiar.

This paper describes a new class of quantitative
Quantum Physics Jeopardy problems that exploit a zero-curvature
($E=0$) energy eigenstate that occurs when suitably chosen Dirac
delta functions are added to an infinite well. These special
configurations are often overlooked,\cite{footnote} but they provide
additional exactly-solvable problems in quantum
mechanics.

\section{Infinite Well with Delta Functions}
In the infinite square well the $E=0$ energy eigenstate is rejected
by most textbooks on the grounds that the general form for energy
eigenstates, $A\sin(kx+\phi)$, does not yield a physical solution
when $k=0$. As pointed out in Ref.~\onlinecite{isw_mistake} the $E=0$
energy eigenstate of the Schr\"{o}dinger equation is not a sinusoidal
function. Instead, $E-V(x)$ is zero inside the well and the
time-independent Schr\"{o}dinger equation simplifies to
$d^2\psi(x)/dx^2 =0$, and yields $\psi(x)=Ax+B$, where $A$ and $B$
are constants. In this context the energy eigenstate cannot be
normalized and still satisfy the boundary and continuity conditions
for the infinite square well, and thus the $E=0$ state cannot be an
allowed energy eigenstate.

Although the authors of Ref.~\onlinecite{isw_mistake} used the
zero-curvature case to show that the infinite square well cannot have
a zero-energy eigenstate, cases in which zero-curvature states are
valid are seldom considered,\cite{gilbert,zc_paper} even though, for
example, all of the energy eigenstates for the infinite square well
are linear (namely zero) outside of the well.

A zero-energy eigenstate can occur with the addition of a single
attractive Dirac delta function , $V_1(x)=-\alpha\delta(x)$ with
$\alpha>0$ , at the origin of a symmetric well with infinite walls at
$x=\pm L$. We write the delta function in this way to work with
positive constants and to make explicit the minus sign in $V_1(x)$.
Similar scenarios have been considered\cite{senn_numerical,rick_susy}
and are related to experiments in which a potential energy ``spike''
inserted in a quantum well is modeled by a Dirac delta
function.\cite{marzin,salis} Our approach differs in that we tune
$\alpha$ to be $\frac{2}{L}(\frac{\hbar^2}{2m})$ so that an energy
eigenstate of zero energy arises. The additional potential energy
function splits the well into two regions: region I ($-L<x<0$) and
region II ($0<x<L$). Assuming a zero-energy eigenstate exists,
continuity requires that it be represented as
$\psi_{\textrm{I}}(x)=A(x+L)$, $\psi_{\textrm{II}}(x)=-A(x-L)$, and
zero outside the well. We then ensure that the energy eigenstate has
the proper discontinuity in its slope at the origin due to the Dirac
delta function. In general, when the Dirac delta function occurs at
the position $x_0$, we must have that
\begin{equation}
\psi^\prime(x_{0^+})-\psi^\prime(x_{0^-})=-\alpha(2m/\hbar^2)\psi(x_0).
\label{delta_slope_change}
\end{equation}
It is easy to show that the energy eigenstate and the chosen value of
$\alpha$ satisfy Eq.~(\ref{delta_slope_change}). Figure~1 shows the
eigenstate corresponding to a symmetric infinite square well with
$V_1(x)$ added, $\psi_1(x)$, which is also a limiting case of an
analysis that used supersymmetric quantum mechanics.\cite{rick_susy}
There are an indefinite number of combinations of Dirac delta functions
that when added to the infinite square well result in zero-energy
eigenstates.

\begin{figure}[h]
\epsfig{file=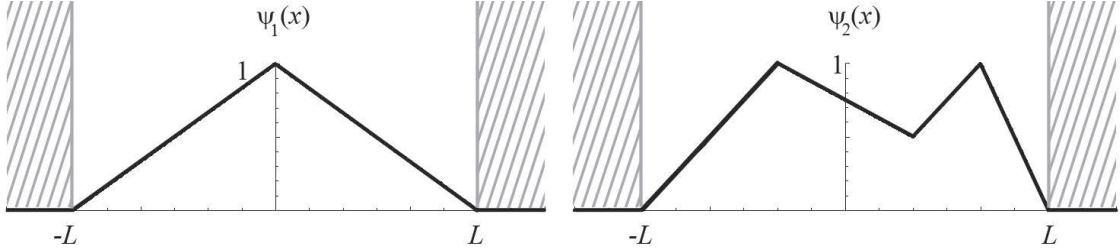,width=15cm,angle=0} \caption{Two
unnormalized zero-energy eigenstates corresponding to two different
sets of Dirac delta function(s) added to a symmetric infinite square
well.}
\end{figure}

We can also proceed in the opposite direction as in Jeopardy: write
any piecewise linear (single-valued) energy eigenstate that vanishes
at $\pm L$ and does \textit{not} vanish at a kink and determine the
Dirac delta function potential(s) that must be added to the infinite
square well. Consider the M-shaped, zero-energy eigenstate
$\psi_2(x)$ in Fig.~1 and determine $V_{2}$. A quantitative result
is possible from direct measurement of the energy eigenstate slopes
at the kinks and their positions. Because
Eq.~(\ref{delta_slope_change}) is independent of an overall
multiplicative factor in $\psi(x)$, we can even begin with an
unnormalized energy eigenstate. The answer turns out to be
$V_{2}=-\frac{9}{4L}(\frac{\hbar^2}{2m})[\delta(x+L/3)
-2\delta(x-L/3)+2\delta(x-2L/3)]$.

We have created a worksheet template for these Jeopardy
exercises.\cite{worksheet,EPAPS} Students can be given individual
problems by replacing the figure in the worksheet with another
drawing of a piecewise linear and single-valued energy eigenstate
that vanishes at the infinite walls and does not vanish at
a kink; the eigenstate can even be drawn with a ruler and graph
paper.

A positive side effect of these Jeopardy problems is that they
further illustrate how energy eigenstates with obvious kinks can be
valid states. This fact is not discussed in most introductory texts
on quantum mechanics. Energy eigenstates must be smooth only if
their corresponding potential energy function is well behaved. Only
infinite walls (such as in the boundaries of the infinite square
well) and Dirac delta functions behave badly enough to generate
kinks in energy eigenstates. It can also be shown that these states
exhibit kinetic/potential energy sharing, by calculating $\langle
\hat{T} \rangle$ and $\langle V_\delta \rangle$ which yields an
exact cancellation so that $\langle E \rangle =0$ for these states
(as expected). This kinetic/potential analysis provides another
illustration of the quantum-mechanical virial
theorem.\cite{zc_paper}

\section{Conclusion}
Zero-energy eigenstates extend the standard treatment of the infinite
square well and other piecewise-constant potential energy
wells.\cite{lin_smit} Although these states seem like an intuitively
natural interpolation between the much more commonly discussed
oscillatory and tunneling solutions, the unfamiliar mathematical form
of the one-dimensional Schr\"{o}dinger equation for situations where
$E-V(x)$ is zero over an extended region of space catches many
students by surprise.\cite{gilbert}

\begin{acknowledgments}
We would like to thank Gary White for useful conversations regarding
this work. LPG and MB were supported in part by a Research
Corporation Cottrell College Science Award (CC5470) and MB was also
supported by the National Science Foundation (DUE-0126439 and
DUE-0442581).
\end{acknowledgments}

\newpage

\begin{figure}
\epsfig{file=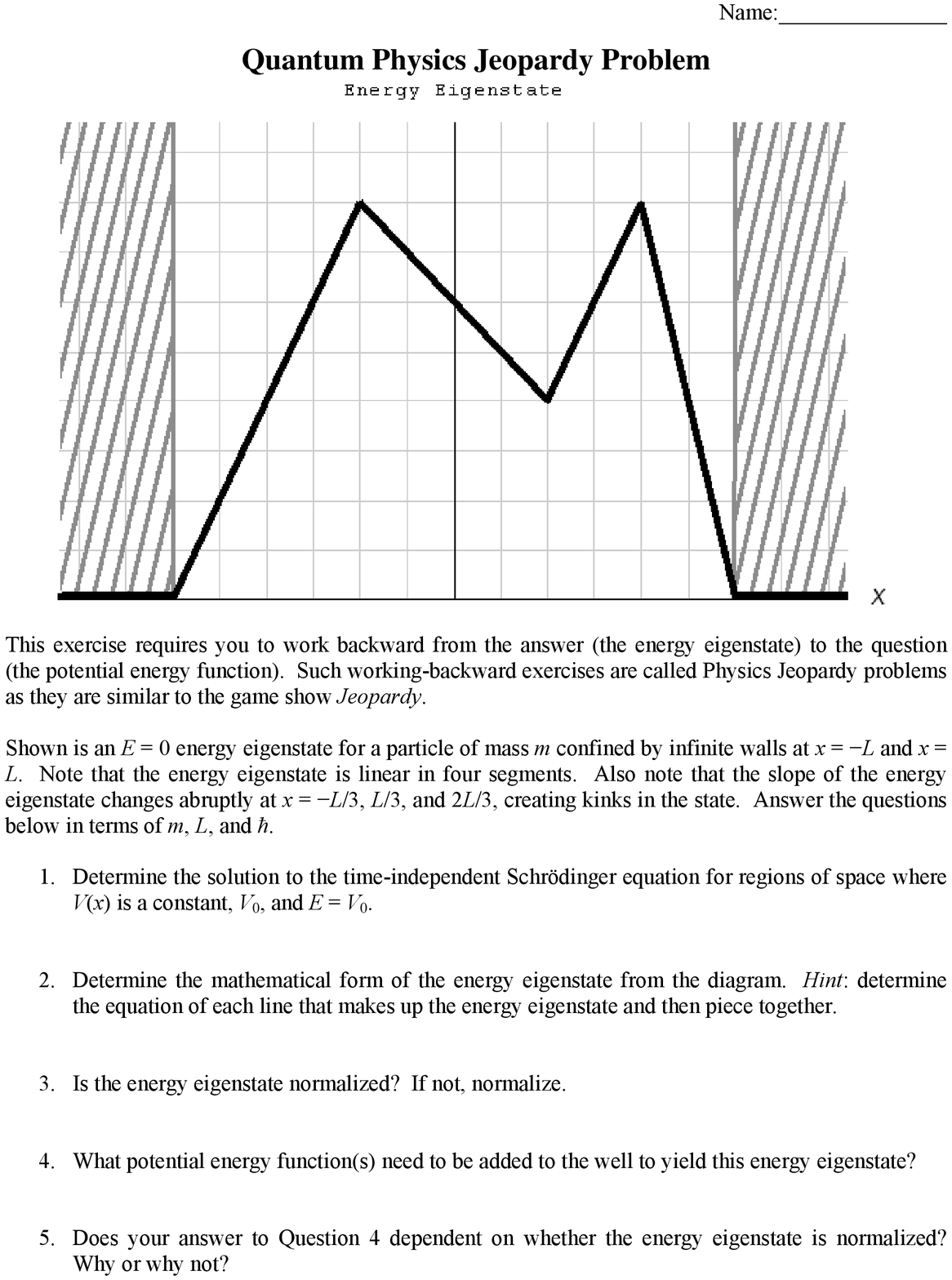,width=16cm,angle=0}
\end{figure}

\begin{figure}
\epsfig{file=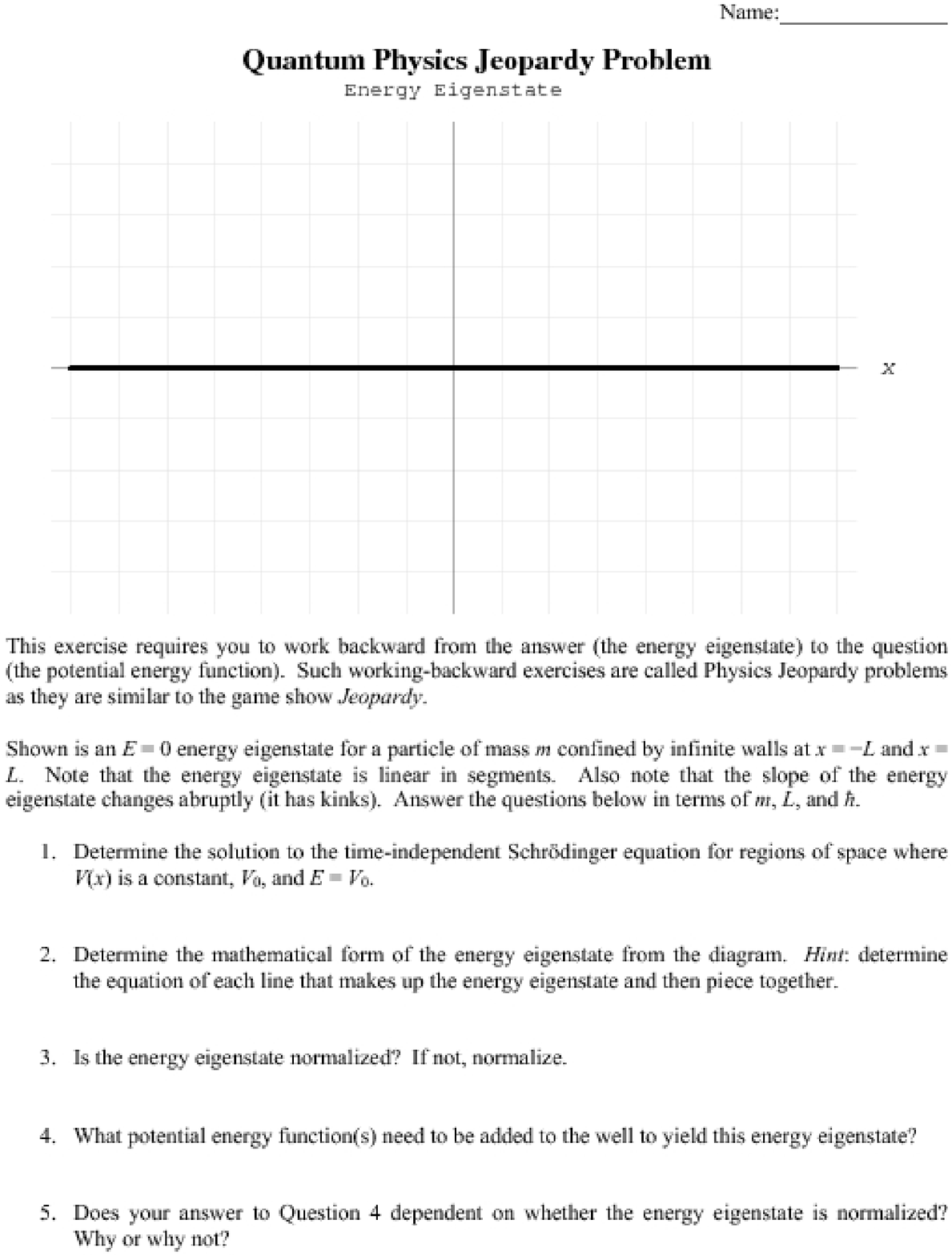,width=16cm,angle=0}
\end{figure}

\end{document}